\documentclass[mathleft]{an}
\usepackage{graphicx}
\usepackage{times}
\usepackage[usenames]{color}

\begin{document}

\Pagespan{789}{}
\Yearpublication{2010}%
\Yearsubmission{2010}%
\Month{11}%
\Volume{999}%
\Issue{88}%

\title{Ground-based follow-up in relation to {\it Kepler} Asteroseismic Investigation\thanks{Based on observations made with the Isaac Newton Telescope and William Herschel Telescope operated by the Isaac Newton Group, with the Nordic Optical Telescope, operated jointly by Denmark, Finland, Iceland, Norway, and Sweden, with the Italian Telescopio Nazionale Galileo (TNG) operated  by the Fundaci\'on Galileo Galilei of the INAF (Istituto Nazionale di Astrofisica), and with the Mercator telescope, operated by the Flemish Community,  all on the island of La Palma at the Spanish Observatorio del Roque de los Muchachos of the Instituto de Astrof\'{\i}sica de Canarias (IAC). Based on observations made with the IAC-80 operated on the  island of Tenerife by the IAC  at the Spanish Observatorio del Teide. Also based on observations taken at the observatories of Sierra Nevada, San Pedro M\'artir, Vienna, Xinglong, Apache Point, Lulin, Tautenburg, McDonald, Skinakas, Pic du Midi, Mauna Kea, Steward Observatory,  Mt. Wilson, Bia\l{}k\'ow Observatory of the Wroc\l{}aw University, Piszk\'estet\H o Mountain Station, and Observatoire de Haute Provence. Based on spectra taken at the Loiano (INAF-OA Bologna), Serra La Nave (INAF - OA Catania) and Asiago (INAF - OA Padova) observatories. Also based on observations collected at the Centro Astron\'omico Hispano Alem\'an (CAHA) at Calar Alto, operated jointly by the Max-Planck-Institut f\"ur Astronomie and the Instituto de Astrof\'{\i}sica de Andaluc\'{\i}a (CSIC). We acknowledge with thanks the variable star observations from the AAVSO International Database contributed by observers worldwide and used in this research.}}

\author{K. Uytterhoeven\inst{1}\fnmsep\thanks{\email{katrien.uytterhoeven@cea.fr}}, M. Briquet\inst{2}, H. Bruntt\inst{3}, P. De Cat\inst{4}, S. Frandsen\inst{5}, J. Guti\'errez-Soto\inst{6}, L. Kiss\inst{7,8}, D.W. Kurtz\inst{9}, M. Marconi\inst{10}, J. Molenda-\.{Z}akowicz\inst{11}, R. {\O}stensen\inst{2}, S. Randall\inst{12}, J. Southworth\inst{13}, R. Szab\'o\inst{7}, \& KASC Working Groups on ground-based observations
}
\titlerunning{Ground-based follow-up of {\it Kepler}}
\authorrunning{K. Uytterhoeven et al.}
\institute{
Lab.\, AIM, CEA/DSM-CNRS-Universit\'e Paris Diderot; CEA, IRFU, SAp, Saclay, 91191, Gif-sur-Yvette, France
\and 
Instituut voor Sterrenkunde, KULeuven, Celestijnenlaan 200D, 3001 Leuven, Belgium
\and
LESIA, Observatoire de Paris-Meudon, 92195 Meudon, France
\and 
Royal Observatory of Belgium, Ringlaan 3, 1180 Brussel, Belgium
\and
Department of Physics and Astronomy, Aarhus University, 8000 Aarhus C, Denmark
\and
Instituto de Astrof\'{\i}sica de Andaluc\'{\i}a (CSIC), Apartado 3004, 18080 Granada, Spain
\and
Konkoly Observatory of the Hungarian Academy of Sciences, 1121 Budapest, Hungary
\and
Sydney Institute for Astrophysics, School of Physics, University of Sydney, Australia
\and
Jeremiah Horrocks Institute of Astrophysics, University of Central Lancashire, Preston PR1 2HE, UK
\and
INAF − Osservatorio Astronomico di Capodimonte, Via Moiariello 16, 80131 Napoli, Italy
\and
Instytut Astronomiczny, Uniwersytet Wroc\l{}awski, Kopernika 11, 51-622
Wroc\l{}aw, Poland
\and
European Southern Observatory, Karl-Schwarzschild-Str. 2, 85748 Garching bei München, Germany
\and 
Department of Physics, University of Warwick, Coventry CV4 7AL, UK
}

\received{01 April 2010}
\accepted{--}
\publonline{later}


\abstract{The {\it Kepler} space mission, successfully launched in March 2009, is providing continuous and high-precision photometry of thousands of stars simultaneously. The uninterrupted time-series of stars of all known pulsation types are a precious source for asteroseismic studies. The {\it Kepler} data do not provide information on the physical parameters, such as $T_{\rm eff}$,  $\log g$, metallicity, and $v \sin i$, which are crucial for successful asteroseismic modelling. Additional ground-based time-series data are needed to characterize mode parameters in several types of pulsating stars. Therefore, ground-based multi-colour photometry and mid/high-resolution spectroscopy are needed to complement the space data. We present ground-based activities within KASC on selected asteroseismic {\it Kepler} targets of several pulsation types.}

\maketitle

\section{Introduction}
The {\it Kepler} satellite (Borucki et al. 1997), launched in March 2009, is collecting light curves with an  unprecedented long time span of 3.5 years and a precision at the level of several ppm (Gilliland et al. 2010) for thousands of stars simultaneously.  Of all {\it Kepler} targets, more than 5000 stars have been selected  as potential targets for seismic studies by the {\it Kepler} Asteroseismic Science Consortium, KASC\footnote{http://astro.phys.au.dk/KASC}. The selection of asteroseismic targets is one of the three basic aims of the KAI ({\it Kepler} Asteroseismic Investigation). The other two are the asteroseismic characterization of planet hosting stars, e.g. derivation of accurate ages, masses, and radii  (e.g. Stello et al. 2009), and the comparison of general stellar properties of Main Sequence stars with those of evolved stars. 

To fully exploit the excellent {\it Kepler} light curves for asteroseismic means and to reach the science goals of the KAI, additional information from ground-based multi-colour photometry and high-resolution spectroscopy is indispensable (see, e.g., Uytterhoeven et al. 2009; Uytterhoeven 2009). The KASC subWorking Groups on ground-based observations (GBOsWG) take care of the organisation of  ground-based observations in support of the {\it Kepler} space data. In this paper we outline the importance of these efforts and we present an overview of the ground-based activities carried out within KASC to date.

\section{{\it Kepler} and the need for ground-based observations}
The need for ground-based {\it Kepler} support data is motivated by two main objectives: 1) the characterization of all {\it Kepler} targets in terms of fundamental stellar parameters, 2) the identification of mode parameters from  multi-colour and spectral time-series observations for selected pulsators. 

\subsection{Characterization of asteroseismic targets}
Crucial for a successful asteroseismic modelling are strong constraints on basic stellar parameters, such as effective temperature ($T_{\rm eff}$), gravity ($\log g$), metallicity ($[M/H]$), and the projected rotational velocity ($v \sin i$). The {\it Kepler} `white light' space data do not provide information on any of these basic stellar parameters, nor on the spectral types of the targets. The latter are essential to correctly classify the stars in terms of pulsation class and evolutionary status. For all these reasons, additional input is needed. 

A first effort to compile a catalogue of stellar parameters for all {\it Kepler} stars, derived from ground-based photometry, has been undertaken by Latham et al. (2005), resulting in the  {\it Kepler} Input Catalogue (KIC). The KIC values are derived from Sloan photometry with the purpose of distinguishing giant stars from dwarfs and are optimised for the characterization of F, G, K, and M dwarfs. Unfortunately, the accuracy of the derived parameters is not good enough for the characterization of more massive or peculiar stars and is too low for seismic modelling. Consequently, more accurate values are needed.

 This problem motivated the GBOsWG to organise a systematic characterization of {\it all} asteroseismic KASC targets. The first aim is to gather for all targets at least one high-resolution spectrum, and preferably two spectra, to derive the spectral type, $\log g$, $T_{\rm eff}$, $[M/H]$, $v \sin i$ (Sousa et al. 2008; Frasca et al. 2006; Niemczura et al. 2009), and to extract information on the stellar environment and possible binarity. The second aim is to obtain more accurate photometric data, in Str\"omgren or Johnson passbands, to derive accurate values of  $\log g$, $T_{\rm eff}$, absolute magnitude, $[M/H]$, and  interstellar reddening (e.g. Bruntt et al. 2004). 

\subsection{Ground-based time-series}
For stars more massive than the Sun and for eclipsing binaries, ground-based time-series are complementary to the {\it Kepler} space light curves. For solar-like oscillators, mode identification relies on the regularity of the frequency pattern in the power spectrum (e.g. Mathur et al. 2010). This method is not directly applicable to larger amplitude pulsators, for which a combination of non-linear effects, rotation, and convection selects the observed modes in a way that is not yet fully understood (e.g. Miglio et al. 2008; Su\'arez et al. 2005, 2009; Degroote et al. 2010). For these targets, the identification of the mode parameters associated with the frequencies observed by {\it Kepler} requires ground-based multi-colour and spectral time-series analysis (e.g. Briquet et al. 2009; Uytterhoeven et al. 2008; Poretti et al. 2009). 
Information on the degree $\ell$ of the modes can be derived from the amplitudes and phases in the different photometric passbands (e.g. Bruntt et al. 2007; De Cat et al. 2007; Handler et al. 2006). The values of both $\ell$ and azimuthal number $m$ can be extracted from the line-profile variations visible in time-series of high-resolution spectra (e.g. Uytterhoeven et al. 2004; Zima et al. 2006; Briquet et al. 2005; Telting et al. 2010). Consequently, dedicated ground-based time-series are needed to complement the accurately derived frequencies (from {\it Kepler} data) with mode parameters and to perform a subsequent seismic modelling.

Multi-epoch spectroscopy is also precious in the study of binaries with a pulsating component. First, through the spectra it is possible to derive constraints on the component masses (e.g. Tango et al. 2006; Vu\v{c}kovic et al. 2007; Creevey et al. 2009; Desmet et al. 2010), which is valuable information for seismic models and models on stellar structure and (binary) evolution.  Second, in the case of double-lined spectroscopic  binaries it is  possible to disentangle the component spectra (Harmanec et al. 2004) and to study the line-profile variability of the disentangled components in full detail (Uytterhoeven et al. 2005). The disentangling of the component contributions is a real advantage over {\it Kepler}'s combined light.

\section{Observational challenges}
The ground-based follow-up of thousands of KASC targets is obviously very challenging and requires  a long-term pro\-ject and a large scale collaboration involving tens of telescopes and instruments. Some observational facts:

$\bullet$  With a 2m-class telescope only about 14 targets can be observed per night, in either multi-colour filters or with spectroscopic instruments. For smaller telescopes the number of observable targets per night is even fewer. To observe all 5000+ targets at least once, more than 360 {\it clear} observing nights are needed. In case we want to obtain a spectrum {\it and} colour information for each star, the total amount of observing time is doubled, i.e. more than 720 nights!

$\bullet$ The {\it Kepler} targets are relatively faint, $V$ magnitudes range between 7 and 16, which makes spectroscopic monitoring very challenging.  For spectral characterization a signal-to-noise ratio (SNR) of at least 50 is needed. A time-series and abundance analysis require high-resolution ($R > 40\,000$) spectra with SNR\,$>$\,80, or preferably SNR\,$>$\,100. To reach the proposed SNR for the fainter targets is not trivial. For example, for a high-resolution spectrograph on a 2m-class telescope the integration times to reach a SNR value of 50, 80, and 100, are 10\,min, 20\,min, and 45\,min, respectively for a $V=10$ star. The corresponding exposure times for a $V=11$ star are tripled. To reach SNR\,$>$\,100 is hence only feasible when several shorter exposures are combined together. These values teach us the following: 
1) It is very time consuming to obtain spectra of sufficient quality for the fainter KASC targets. 
2) Because of the trade-off between pulsation period and exposure time, spectroscopic time-series are only feasible for bright targets ($V<9.5$). 
3) We preferably need larger telescopes ($>$2m-class mirror), equipped with a high-resolution spectrograph ($R>40\,000$). Unfortunately, such telescope -- instrument combinations are scarce.

\section{Observations}
\begin{figure}
\centering
\includegraphics[width=0.35\textwidth]{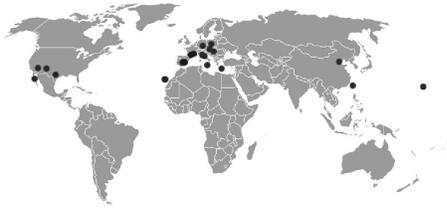}
\caption{World map with the observatories involved in the KASC ground-based observations indicated by dark grey dots.}
\label{Map}
\end{figure}

To date, a total of more than 530 observing nights has been awarded to the GBOsWG on 36 different instruments at 31 telescopes  on 23 observatories in 12 countries. Figure~\ref{Map} gives an overview of the observatories that are, to date, included in the KASC ground-based observational project.  Here we present the highlights of the observations on target characterization ($>$278 nights, 26 different instruments on 17 observatories) and time-series analysis ($>$256 nights, 15 different telescopes on 13 observatories). For a detailed overview of the awarded observing time we refer to Uytterhoeven et al. (2010).

{\bf $\bullet$ Characterization of Solar-like stars: }
Since 2004, a project has been running to characterize KASC solar-like stars (Molenda-\.{Z}akowicz et al. 2007, 2008, 2009b). In total, 104 nights  of spectroscopic and 10 nights of photometric observations have been performed with the following instruments: SARG@TNG, La Palma (E), FRESCO and the  single-channel photometer@0.91m, Ca\-ta\-nia (I). Also, an interferometric project is ongoing with PAVO@CHARA at Mt Wilson Observatory (USA) to measure angular diameters for some of the brighest Kepler targets. In addition, observations are scheduled with FIES@NOT, La Palma (E), and a total of 30 hours has been awarded with ESPaDOnS\-@CFHT, Mauna Kea (USA), and  NARVAL@TBL, Pic Du Midi (F), in high-resolution ($R=80\,000$) mode. 

{\bf $\bullet$ Characterization of several pulsation classes:}
Several observing proposals for the characterization of a mixture of pulsation classes have been and are being prepared.
For the semester running from October 2009 to August 2010 a total of 44 nights of photometric observations and 45 nights of spectroscopy, with 13 different instruments, has been a\-war\-ded for the characterization of $\gamma$\,Dor, $\delta$\,Sct, $\beta$\,Cep, Be, solar-like, roAp, and Slowly Pulsating B (SPB) stars,  and stars in clusters (see Table 1 in Uytterhoeven et al. 2010). The photometric instruments involved are: BUSCA@2.2m, Calar Alto (D);  Str\"omgren photometer@1.5m,  San Pedro M\'artir (MX); Str\"omgren photometer@0.90m Sierra Nevada (E); CCD@1.0m RRC, Piszk\'estet\H o (H); CAMELOT@IAC-80, Teide (E); WFC@INT, La Palma (E). The spectroscopic instruments are: Coud\'e@2m, Tautenburg (D); cs23@2.7m, McDonald (USA); BFOSC@1.52m, Asiago (I); FIES@NOT and HERMES@Mercator, La Palma (E); FRESCO@0.91m, Catania (I); spectrograph@2.12m,  San Pedro M\'artir (MX);   SOPHIE@1.92m, Haute Provence (F).

 In addition, an ambitious proposal to observe 95\% of all KASC asteroseismic targets with the multi-fiber, multi-object spectrograph LAMOST@4m telescope at Xinglong observatory (CN)  has been submitted.

{\bf $\bullet$ Characterization of K giants:}
 A total of 13 nights has been awarded since 2009 for the characterization of K giants with FIES@NOT, La Palma (E). High-resolution spectra of giant stars in the cluster NGC6811 will be obtained with ESPaDOnS@CFHT, Mauna Kea (USA). 

{\bf $\bullet$ Characterization of compact sdOB pulsators:}
 In total, 28 nights of spectroscopic observations have been carried out/are scheduled for the characterization of compact sdOB pulsators, with three different instruments: IDS@INT and  ISIS@WHT, La Palma, (E), and the B\&C spectrograph at the 2.3m Bok telescope, Steward Observatory (USA). Additional spectra have been collected with FIES@NOT, La Palma, (E).

{\bf $\bullet$ Characterization of $\delta$ Sct stars:}
Several Italian instruments and telescopes have been used to obtain spectra for 19 KASC $\delta$\,Sct stars in 2009:  BFOSC@1.52m, Loiano (I), FRESCO@0.91m, Catania (I), SARG@TNG, La Palma (E), and  AFOSC@1.82m, Asiago (I). Physical parameters of these targets have been derived by Catanzaro et al. (2010).

{\bf $\bullet$ Characterization of Be stars:}
  Two KASC Be candidates were observed in 2008 and 2009 with different spectroscopic and photometric instruments, during a total of eight nights, with the aim to characterize and describe their Be nature.  At least six more observing nights are awarded in 2010. The instruments and telescopes involved are: 1.3m telescope, Skinakas (GR), ALBIREO@1.52m and photometer@0.9m,  Sierra Nevada (E), and  ALFOSC@NOT, La Pal\-ma (E). 

{\bf $\bullet$ Spectropolarimetric characterization:} 
To investigate magnetic signatures, spectropolarimetric observations are planned for selected solar-like, RR Lyr, $\delta$\,Sct, and Be stars with ESPaDOnS@CFHT, Mauna Kea (USA). 

{\bf $\bullet$ Time-series of pulsators in clusters:}
A large photometric multi-site campaign was carried out in 2009 on the cluster NGC 6866 and a similar observational effort is being organised for the cluster NGC 6811 in 2010.  The cluster NGC 6866  is known to host at least three $\delta$\,Sct and two $\gamma$\,Dor candidates (Molenda-\.Zakowicz et al. 2009a), and there are 12 known $\delta$\,Sct stars in NGC 6811 (Luo et al. 2009). The goal of the project is to perform mode iden\-ti\-fi\-ca\-tion with multi-band photometry. We refer to Table\,2 in Uytterhoeven et al. (2010) for an overview of the involved instruments and telescopes. The following nine observatories are included in the multi-site campaign: Apache Point (USA), Teide (E), Sierra Nevada (E), Catania (I), Loiano (I), Vienna (A), Bia\l{}k\'ow (PL), Piszk\'estet\H o (H), and Xinglong (CN).

{\bf $\bullet$ Time-series of RR Lyr stars and Cepheids:}
Multi-colour time-series of RR Lyr stars and Cepheids will be obtained during a total of 21 nights in 2010 with SLT@0.4m and LOT@1.0m at Lulin Observatory in Taiwan.  In addition, follow-up observations of selected promising targets are being performed in the framework of AAVSONet.

{\bf $\bullet$ Time-series of SPBs and hybrid $\gamma$\,Dor/$\delta$\,Sct pulsators:}
For the multi-colour photometric monitoring of hybrid $\gamma$\,Dor/$\delta$\,Sct pulsators 14 nights have been awarded in 2010 with the 0.90m telescope at Sierra Nevada observatory (E). In addition, seven and eleven nights have been allocated for the spectroscopic monitoring of selected SPBs and $\gamma$\,Dor hybrids with CE@2.1m at McDonald observatory (USA), and HERMES@Mercator, La Palma (E), respectively.

\section{Future plans and call for your contribution}
The ground-based observational and organisational efforts of the KASC GBOsWG have been very successful so far, with already more than 530 observing nights awarded. This very important task of supporting {\it Kepler} from the ground revives the use of small/mid-sized telescopes, which is a significant benefit for all the national observatories involved.  Given the large scale of the project, and the tremendous amount of observing time needed, the KASC is putting heavy pressure on ground-based telescopes in the Northern hemisphere, especially on the ones equipped with a high-re\-so\-lu\-tion spectrograph. Therefore, any help is welcome!  Please contact us if you have access to (further) telescopes and want to join the project. Moreover, we also need more observers and  helping hands in the reduction and analysis of the flood of ground-based data, to be able to timely provide analysis results. The large amount of data also calls for a large database, accessible to all observers involved to share information,  to keep track of the observed targets, and to avoid duplication. The seismic and non-seismic information provided by the GBOsWG, in combination with the excellent {\it Kepler} space photometry, promise a bright future for asteroseismology.

\begin{acknowledgements}
We thank all KASC members that are contributing to {\it Kepler} ground-based observations. MB is Postdoctoral Fellow of the Fund for Scientific Research, Flanders. This work was supported by  MNiSW grant N203 014 31/2650, and by the National Office for Research and Technology through the Hungarian Space Office Grant No. URK09350 and the `Lend\"ulet' program of the Hungarian Academy of Sciences.
\end{acknowledgements}

\end{document}